\newcommand{\ifLong}[1]{#1}
\newcommand{\ifShort}[1]{}
\newcommand{\ifShortVspace}[1]{}
\renewcommand{\ifLong}[1]{}
\renewcommand{\ifShort}[1]{#1}
\renewcommand{\ifShortVspace}[1]{\vspace{#1}}

\documentclass[runningheads,final]{llncs}
\usepackage[utf8]{inputenc}
\usepackage[T1]{fontenc}
\usepackage{graphicx}
\graphicspath{ {./images/} }

\usepackage[english]{babel}
\usepackage{subcaption}
\usepackage{url}
\usepackage{colortbl}
\usepackage{amssymb}
\usepackage{stmaryrd}
\usepackage{amsmath}
\usepackage{mathrsfs}
\usepackage{amssymb}
\usepackage[left]{lineno}
\usepackage{xfrac}
\usepackage{nicefrac}
\usepackage[textsize=scriptsize,backgroundcolor=yellow!40,obeyFinal]{todonotes}
\usepackage[nonumberlist,acronym,sanitize=none]{glossaries}
\glsdisablehyper
\usepackage{comment}
\usepackage[pdftex, colorlinks=true, hyperfootnotes=true, hyperindex=true,
            plainpages=false, pagebackref=false, pdfpagelabels=true, pdfstartview=FitH,
            linkcolor=blue, citecolor=blue, urlcolor=blue,
            bookmarks, bookmarksopen, bookmarksdepth=3]{hyperref}
\usepackage[capitalise,nameinlink]{cleveref}
\crefname{algocf}{alg.}{algs.}
\Crefname{algocf}{Algorithm}{Algorithms}
\crefname{section}{Sect.}{Sects.}
\Crefname{section}{Section}{Sections}
\usepackage{tkz-base}
\usetikzlibrary{decorations.pathmorphing,trees,snakes,arrows,shapes,automata,petri}
\usepackage{paralist}
\usepackage{multicol}
\usepackage{booktabs}
\usepackage{enumitem}
\usepackage{multirow}
\usepackage{rotating}
\usepackage{etaremune}
\usepackage{marginnote}
\usepackage{mathtools}
\usepackage{mathabx}
\usepackage{adjustbox}
\usepackage{ifthen}
\usepackage[normalem]{ulem}
\usepackage{lineno}
\usepackage{soul}
\usepackage{floatrow}
\usepackage{listings}
\crefname{lstlisting}{Listing}{Listings}
\usepackage{lipsum}
\usepackage{makecell}
\usepackage{diagbox}
\usepackage[scientific-notation=false]{siunitx}
\usepackage{footmisc}
\usepackage{xspace}
\usepackage{ifdraft}
\usepackage{wrapfig}%
\usepackage{times}

\usepackage[subtle]{savetrees}
\newacronym[\glslongpluralkey={Distributed Ledger Technologies}]{dlt}{DLT}{Distributed Ledger Technology}
\newcommand{\DLT}[0] {\Gls{dlt}\xspace}
\newcommand{\DLTs}[0] {\Glspl{dlt}\xspace}

\newacronym{ipfs}{IPFS}{InterPlanetary File System}
\newcommand{\IPFS}[0] {\Gls{ipfs}\xspace}

\newacronym{p2p}{P2P}{peer-to-peer}
\newcommand{\PtoP}[0] {\Gls{p2p}\xspace}

\newacronym{abe}{ABE}{Attribute-Based Encryption}
\newcommand{\ABE}[0] {\Gls{abe}\xspace}

\newacronym{maabe}{MA-ABE}{Multi-Authority Attribute-Based Encryption}
\newcommand{\MAABE}[0] {\Gls{maabe}\xspace}

\newacronym{cpabe}{CP-ABE}{Ciphertext-Policy Attribute-Based Encryption}
\newcommand{\CPABE}[0] {\Gls{cpabe}\xspace}

\newacronym{ssl}{SSL}{Secure Sockets Layer}

\newacronym{dht}{DHT}{Distributed Hash Table}
\newcommand{\DHT}[0] {\Gls{dht}\xspace}

\newacronym{mpc}{MPC}{Multi-party Computation}
\newcommand{\MPC}[0] {\Gls{mpc}\xspace}

\newglossaryentry{box}{ %
  name={Box},
  description={authority intialisation storage box},
  first={\glsentrydesc{box} (henceforth, \glsentrytext{Box} for short)},
  plural={Boxes},
  firstplural={\glsentrydesc{box}es (henceforth, \glsentryplural{box} for short)}
}

\newglossaryentry{rloc}{ %
	name={resource locator},
	description={a generic term to denote content-based links obtained via hashing (e.g., the IPFS link)},
}
\def\rloc{\gls{rloc}\xspace}

\newacronym{cpmaabe}{CP-MA-ABE}{Ciphertext-Policy Multi-Authority Attribute-Based Encryption}

\def\DOwner{Data Owner\xspace}
\def\AttCert{Attribute Certifier\xspace}
\def\Reader{Reader\xspace}
\def\SC{Smart Contract\xspace}
\def\MeC{Message Contract\xspace}
\def\AtC{Attribute Certifier Contract\xspace}
\def\AutC{Authority Contract\xspace}
\def\Auth{Authority\xspace}
\def\Auths{Authorities\xspace}
\def\DStore{Data Store\xspace}

\newacronym{dk}{\textsl{dk}}{decryption key}
\newacronym{fdk}{\textsl{fdk}}{final decryption key}
\def\dk{\gls{dk}\xspace}
\def\fdk{\gls{fdk}\xspace}
\newacronym[\glslongpluralkey={Business Processes}]{bp}{BP}{Business Process}
\newacronym{bpi}{BPI}{Business Process Intelligence}
\newacronym{bpm}{BPM}{Business Process Management}
\newacronym{bpms}{BPMS}{Business Process Management System}
\newacronym{bpmn}{BPMN}{Business Process Model and Notation}
\newacronym{cpn}{CPN}{colored Petri net}
\newacronym{kpi}{KPI}{Key Performance Indicator}
\newacronym{ocbc}{OCBC}{Object-centric Behavioral Constraints}
\newacronym{soa}{SOA}{Service-Oriented Architecture}
\newacronym{pn}{PN}{Petri net}
\newacronym{wf}{WF}{workflow}
\newacronym{wfms}{WfMS}{Workflow Management System}
\newacronym{xes}{XES}{eXtensible Event Stream}
\newacronym{yawl}{YAWL}{Yet Another Workflow Language}
\newglossaryentry{task}{%
	name={task},description={the non-divisible, elementary activity}}

\newglossaryentry{promod}{%
	name={process model},description={the model of a process}
}
\def\LogAlph {\ensuremath{\Sigma}}
\newglossaryentry{logalph}{
	name={log alphabet},description={the process alphabet, as reflected in a log},%
	symbol={\LogAlph}}
\def\Evt {\ensuremath{e}}
\newglossaryentry{evt}{
	name={event},description={a record of an instantaneous fact during the process enactment},%
	symbol={\Evt}}
\def\Trc { \ensuremath{\tau} }

\newglossaryentry{trace}{
	name={trace},description={a sequence of \glsplural{evt}},%
	symbol={\Trc}}
\def\EvtLog {\ensuremath{L}}

\newglossaryentry{evtlog}{
	name={event log},description={a collection of \glstext{evttrace}s},%
	symbol={\EvtLog}}

\renewcommand\tabcolsep{5pt}
\newcolumntype{d}{>{\columncolor{gray!10}}c}
\newcolumntype{m}{>{\columncolor{gray!10}}l}
\newfloatcommand{capbtabbox}{table}[][\FBwidth]
\newenvironment{iiilist}%
{\begin{inparaenum}[\itshape(i)\upshape]}%
{\end{inparaenum}}

\def\GoodExampleMark{\checkmark}
\def\BadExampleMark{$\times$}

\RequirePackage{xparse}
\NewDocumentEnvironment{AuthNote}{+o+o}{%
	\IfValueT{#2}{\marginnote{\scriptsize{#2}}}%
	\begin{scriptsize}
		\colorbox{gray}%
		{\color{white} Note\IfValueT{#1}{ (#1)}:}%
		\quad%
		\color{brown}
}{%
	\normalcolor
	\end{scriptsize}
}

\RequirePackage{pifont}
\def\CircOne {\ding{192}}
\def\CircTwo {\ding{193}}
\def\CircThree {\ding{194}}
\def\CircFour {\ding{195}}

\RequirePackage{lipsum}
\newcommand{\LipsumGray}[1][]{{\color{gray}\ifthenelse{\equal{#1}{}}{\lipsum}{\lipsum[#1]}}}

\RequirePackage{siunitx}
\newcolumntype{D}[1]{S[
	table-omit-exponent,
	round-mode=places,
	round-integer-to-decimal,
	round-precision={#1}]} 

\newcommand{\ennote}[1]{\ifoptionfinal{#1}{\textit{\textcolor{orange}{#1 --EN}}}}

\providecommand{\ie}{{i.e.,}\xspace}

\newcommand{\Req}[1]{\underline{\textbf{R#1}}}
\begin{document}

\title{MARTSIA: Enabling Data Confidentiality \\ for Blockchain-based Process Execution}
\titlerunning{MARTSIA}

\titlerunning{MARTSIA: Data Confidentiality for Process Execution}
%
\author{Edoardo~Marangone\inst{1}
    \and
	Claudio~Di~Ciccio\inst{1}
    \and
	Daniele~Friolo\inst{1}
    \and	\\
        Eugenio~Nerio~Nemmi\inst{1}
    \and
        Daniele~Venturi\inst{1}
    \and
        Ingo~Weber\inst{2}
}

\authorrunning{Marangone, Di Ciccio, Friolo, Nemmi, Venturi, Weber}
%
\institute{Sapienza University of Rome, Rome, Italy\\
	\email{\href{mailto:marangone@di.uniroma1.it;diciccio@di.uniroma1.it;friolo@di.uniroma1.it;nemmi@di.uniroma1.it;venturi@di.uniroma1.it}{\{marangone,diciccio,friolo,nemmi,venturi\}@di.uniroma1.it}}
	\and
	Technical University of Munich, School of CIT, \\ and Fraunhofer Gesellschaft, Munich, Germany,
    \email{\href{mailto:first.last@tum.de}{first.last@tum.de}}
}

\maketitle

\begin{abstract}
	\vspace{-1em}
	Multi-party business processes rely on the collaboration of various players in a decentralized setting. 
Blockchain technology can facilitate the automation of these processes, even in cases where trust among participants is limited.
Transactions are stored in a ledger, a replica of which is retained by every node of the blockchain network. 
The operations saved thereby are thus publicly accessible.
While this enhances transparency, reliability, and persistence, it hinders the utilization of public blockchains for process automation
as it violates typical confidentiality requirements in corporate settings.
In this paper, we propose \emph{MARTSIA}:
A \textit{Multi-Authority Approach to Transaction Systems for Interoperating Applications}.
MARTSIA enables precise control over process data at the level of message parts. 
Based on Multi-Authority Attribute-Based Encryption (MA-ABE), MARTSIA realizes a number of desirable properties, including confidentiality, transparency, and auditability.
We implemented our approach in proof-of-concept prototypes, with which we conduct a case study in the area of supply chain management. 
Also, we show the integration of MARTSIA with a state-of-the-art blockchain-based process execution engine to secure the data flow.
%
\end{abstract}
%
%
\section{Introduction}
\label{sec:intro}
\begin{sloppypar}
Enterprise applications of blockchain technology are gaining popularity because it enables the design and implementation of business processes involving many parties with little mutual trust, among other benefits~\cite{Weber.etal/BPM2016:UntrustedBusinessProcessMonitoringandExecutionUsingBlockchain,Stiehle22SLR}. 
Standard blockchains yield the capability of enabling cooperation between potentially untrusting actors through transparency: relevant data is made available to all participants of a blockchain network, and hence can be verified by anyone, thereby removing the need for trust~\cite{Xu.etal/2019:ArchitectureforBlockchainApplications}.
In combination with the high-integrity permanence of data and non-repudiability of transactions offered by the technology, blockchains can be used to realize trustworthy protocols.
\end{sloppypar}

However, in multi-party business settings with scarce mutual trust, the involved parties typically have a strong need to keep certain data hidden from some of the business partners, and even more so from most other participants in a blockchain network.
In fact, fulfilling security and privacy requirements is a key obstacle when it comes to the adoption and implementation of blockchain technology in general~\cite{Privacy1,Privacy2}.
Corradini et al.~\cite{Corradini.etal/ACMTMIS2022:EngineeringChoreographyBlockchain} confirm the importance of security and privacy considerations for the specific case of process execution on blockchain. 

Simple cryptographic solutions face severe downsides, as discussed in the following.
First, the authors of~\cite{Corradini.etal/ACMTMIS2022:EngineeringChoreographyBlockchain} note that simply encrypting the contents of messages (payload), as previously proposed in the literature, does not guarantee the confidentiality of the information.
Using synchronous encryption requires sharing a decryption key among process participants, and thus does not allow the sender of the data to selectively control access to different parts of a single message. 
Using asynchronous encryption and encrypting a message with the public key of the recipient requires the sender to create multiple copies of each message (one for each intended reader), which means that the sender can send \textit{different} pieces of information to each participant -- i.e., integrity is lost.
Other proposed solutions address the issue via perimeter security: read access to the (relevant parts of) a blockchain is limited, e.g., by using channels on Hyperledger Fabric or similar~\cite[Ch.~2]{Xu.etal/2019:ArchitectureforBlockchainApplications}, or using private blockchains.
However, this approach suffers from the same downsides as the use of synchronous encryption above.
Also, permissioned platforms require the presence of trusted actors with the privileged role of managing information exchange \emph{and} the right to be part of the network.
In summary, most of the previous approaches offer ``all-or-nothing'' access: either all participants in some set can access the information in a message, or they receive only private messages and integrity of the data sent to multiple recipients is lost.
In previous work~\cite{Marangone.etal/BPM2022:CAKE}, we introduced an early approach to control data access at a fine-granular level. However, the  architecture relied on a central node for forging and managing access keys, thus leading to easily foreseeable security issues in case this single component were to be compromised or byzantine. Also, its integration with process management systems was yet to be verified. With the objective of overcoming these limitations, we have revised the entire approach from its foundations and devised the new solution we present here.
\todo{Highlighting the non-trivial delta with previous work.}

In this paper, we propose a Multi-Authority Approach to Transaction Systems for Interoperating Applications (MARTSIA). In MARTSIA, encrypted data is persisted in decentralized storage, which is connected to a public permissionless blockchain system supporting process execution. 
Data owners define access policies to regulate which users are able to view specific parts of the information. 
No central authority can generate decryption keys alone. The encryption and decryption of messages are left to the individual nodes.
To attain the desired characteristics, this approach employs hybrid encryption in combination with the Ciphertext-Policy variant of \acrlong{maabe} (henceforth, \acrshort{maabe} for the sake of conciseness), smart contracts, and \acrfull{ipfs}. 
In our evaluation, we show the integration of our implemented prototype with Caterpillar~\cite{Lopez-Pintado.etal/SPE2019:Caterpillar}, a state-of-the-art process execution engine, to demonstrate how our approach can complement a business process 
management system to secure its data flow.

In the following, \cref{sec:example} presents a running example, to which we will refer throughout the paper, and illustrates the problem we tackle. \Cref{sec:background} outlines the fundamental notions that our solution is based upon. In \cref{sec:approach}, we describe our approach in detail. In \cref{sec:imptes}, we present our proof-of-concept implementation and the results of the experiments we conducted therewith. \Cref{sec:sota} reviews related work before \cref{sec:conclusion} concludes the paper and outlines future works.

\section{Example, problem illustration, and requirements}
\label{sec:example}
\begin{figure}[tb]
  \includegraphics[width=\textwidth]{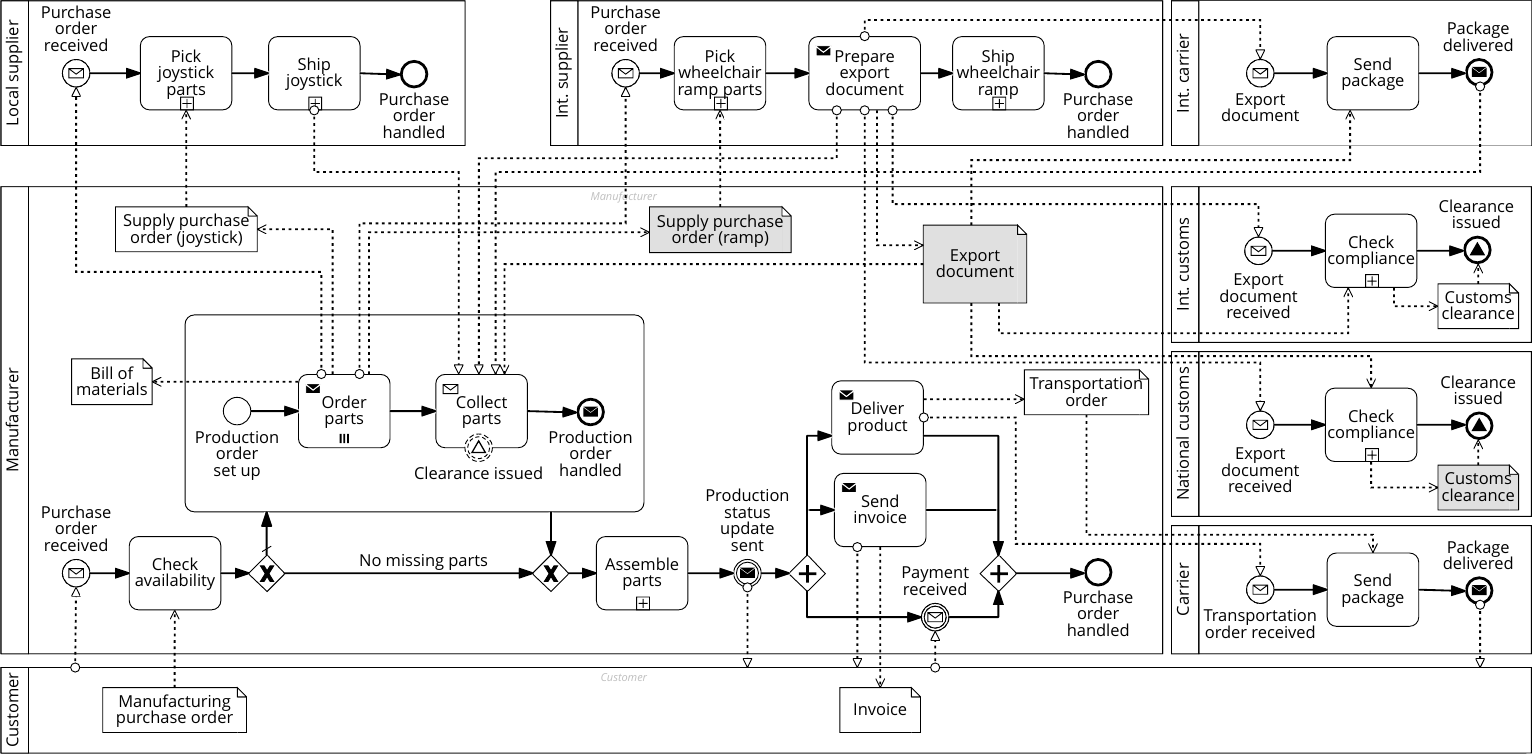}
  \caption{A multi-party process for the assembly of special car parts.}
  \label{fig:example}
\end{figure}

\Cref{fig:example} shows a \gls{bpmn} collaboration diagram illustrating a supply chain in the automotive area: the production of a special car for a person with paraplegia. The two base components for that car are a joystick (to turn, accelerate, brake) and a wheelchair ramp to let the person get into the vehicle.

A new process instance is initiated when a \textit{customer} places an order for a modified car from a \textit{manufacturer}. The \textit{manufacturer} then checks the availability of joystick parts and wheelchair ramps in the warehouse, to order the missing ones from a local \textit{joystick supplier} and an international \textit{wheelchair ramp supplier}, respectively. 
Once all ordered parts have been collected, the suppliers prepare the packages with the products for delivery. The \textit{international customs} 
verifies the \textit{document} of the international supplier and issues custom clearance once the compliance verification is completed successfully. 
The \textit{carrier} of the international supplier checks the documents and delivers the package to the \textit{manufacturer} with an international shipment procedure. Upon receipt of the parts, the \textit{manufacturer} proceeds with the assembly process. After informing the \textit{customer} about the progress of the production process, the \textit{manufacturer} sends an invoice and requests a carrier to deliver the package. The process is completed with the delivery of the ordered product.

\begin{table}[tbp]
	\caption[An excerpt of the messages exchanged]{An excerpt of the information artifacts exchanged in \cref{fig:example} \vspace{-0.75em}}
	\label{tab:messageEncoding:before}
	\resizebox{0.9\columnwidth}{!}{%
\begin{tabular}{|c|c|l|c|}
\hline
   Message
 & Sender
 & Data 
 & Recipients \\
 \hline
 
   \makecell{Supply purchase order (ramp)}
 & Manufacturer
 & 
\begin{lstlisting}[firstline=1,lastline=4]
Company_name:           Alpha
Address:                38, Alpha street
E-mail:                 cpny.alpha@mail.com
Price:                  $5000
\end{lstlisting}
 & \makecell{International supplier}
 \\
 \hline   
 & 
 & 
\begin{lstlisting}[firstline=8,lastline=16]
Manufacturer_company:   Beta
Delivery_address:       82, Beta street
E-mail:                 mnfctr.beta@mail.com
Ramp run:               3
Kickplate:              12
Handrail:               7
Baluster:               30
Guardrail:              25
Amount_paid:            $5000
\end{lstlisting} 
 & \makecell{Manufacturer\\ National customs \\ International customs\\ International carrier}
 \\ 
 \cline{3-4}
   \makecell{Export document}
 & International supplier
 & 
\begin{lstlisting}[firstline=18,lastline=22]
Fundamental_workers_rights:             Ok     
Human_rights:                           Ok
Protection_biodiversity_and_ecosystems: Ok
Protection_water_and_air:               Ok
Combatting_climate_change:              Ok
\end{lstlisting} 
 & \makecell{National customs\\ International customs}
 \\ 
 \cline{3-4} 
 &
 & 
\begin{lstlisting}[firstline=24,lastline=26]
Manufacturer_company:   Beta
Address:                78, Beta street
Order_reference:        26487
\end{lstlisting}
 & \makecell{Manufacturer} 
 \\
 \cline{3-4} 
 & 
 & 
\begin{lstlisting}[firstline=28,lastline=32]
Invoice_ID:             101711
Billing_address:        34, Gamma street
Gross_total:            $5000
Company_VAT:            U12345678
Issue_date:             2022-05-12
\end{lstlisting} 
 & \makecell{Manufacturer\\ National customs\\ International customs}
 \\
 \hline
 
 \makecell{Customs clearance}
 & National customs
 & 
\begin{lstlisting}[firstline=36,lastline=40]
Tax_payment:            confirmed
Conformity_check:       passed
Date:                   2022-05-10
Sender:                 Beta
Receiver:               Alpha
\end{lstlisting} 
 & \makecell{Manufacturer\\ International customs} 
 \\
 \hline

\end{tabular}

	}
\end{table} 

Throughout the paper, we will refer to this scenario as a running example.
In particular, we will focus on the information artifacts in \cref{tab:messageEncoding:before} (marked with a gray background color in \cref{fig:example}), namely:
\begin{inparaenum}[(1)]
	\item\label{item:purchaseorder} the purchase order of the manufacturer, 
	\item\label{item:exportdocument} the export document of the international supplier, and
	\item\label{item:customsclearance} the national customs clearance. 
\end{inparaenum}
The export document encloses multiple records, namely
\begin{inparaenum}[({2}.a)]
	\item\label{item:intshipmentorder} the international shipment order, 
	\item\label{item:csdd} the Corporate Sustainability Due Diligence Directive (CSDD), 
	\item\label{item:orderreference} the reference to the order, and
	\item\label{item:invoice} the invoice.
\end{inparaenum}
These records are meant to be accessed by different players.
The shipment order should only be accessible by the international carrier and the two customs bodies,
the CSDD can only by read by the customs authorities,
the order reference is for the manufacturer,
and the invoice is for the manufacturer and the customs bodies.
Differently from the purchase order and the customs clearance (messages \ref{item:purchaseorder}~and~\ref{item:customsclearance}), the four above entries (\ref{item:exportdocument}.\ref{item:intshipmentorder}~to~\ref{item:exportdocument}.\ref{item:invoice}) are joined in a single document for security reasons: separate messages could be intercepted and altered, replaced or forged individually. Once they are all part of a single entity, every involved actor can validate all the pieces of information.
Ideally, in a distributed fashion every node in the network could be summoned to attest to the integrity of that document.
However, the need for separation of indicated recipients demands that only a selected group of readers be able to interpret the parts that are specifically meant for them (see the rightmost column of \cref{tab:messageEncoding:before}).
In other words, though \emph{visible} for validation, the data artifact should not be interpretable by everyone.
The other actors should attest to data encrypted as in \cref{tab:messageEncoding:after}.
This aspect gives rise to one of the requirements 
we discuss next.

\begin{table}[tb]
	\centering
	\caption[Requirements and corresponding actions in the approach]{Requirements and corresponding actions in the approach \vspace{-0.75em}}
	\label{tab:requirements}
	\resizebox{1.0\columnwidth}{!}{%
	\begin{tabular}{l p{6.5cm} l l l} \toprule
	& \textbf{Requirement} 
	& \textbf{CAKE~\cite{Marangone.etal/BPM2022:CAKE}} 
	& \textbf{MARTSIA} 
	& See 					
	\\ \midrule
	\Req{1} & Access to parts of messages should be controllable in a fine-grained way (attribute level), while integrity is ensured 
	& \GoodExampleMark
	& \GoodExampleMark
	& \Cref{sec:approach:datastructures,sec:imptes}	
	\\
	\Req{2} & Information artifacts should be written in a permanent, tamper-proof and non-repudiable way 
	& \GoodExampleMark
	& \GoodExampleMark
	& \Cref{sec:approach:workflow,sec:imptes}	
	\\
	\Req{3} & The system should be independently auditable with low overhead 
	& \GoodExampleMark
	& \GoodExampleMark
	& \Cref{sec:approach:workflow,sec:imptes}	
	\\
	\Req{4} & The decryption key should only be known to the user who requested it 
	& \BadExampleMark
	& \GoodExampleMark
	& \Cref{sec:approach:workflow,sec:imptes}	
	\\
	\Req{5} & The decryption key should not be generated by a single trusted entity
	& \BadExampleMark
	& \GoodExampleMark
	& \Cref{sec:approach:workflow,sec:imptes}	
	\\
	\Req{6} & The approach should integrate with control-flow management systems
	& \BadExampleMark
	& \GoodExampleMark
	& \Cref{sec:approach:datastructures,sec:imptes}	
\ifLong{%
	\\
	\ennote{\Req{7}} & \ennote{Smart contracts should prevent any disruptive operation unless the majority of the authorities agree.}  
	& \BadExampleMark
	& \GoodExampleMark
	\\
	\ennote{\Req{7.1}} & \ennote{Smart contracts should not be updatable (\ie no changes in any smart contracts bytecode)}  
	& \BadExampleMark
	& \GoodExampleMark
	\\
	\ennote{\Req{7.2}} & \ennote{Smart contracts should only be cancelable if the majority of attribute certifiers agree.}  
	& \BadExampleMark
	& \GoodExampleMark%
}	
	\\
	
	\bottomrule
\end{tabular}

	}
\end{table}
%
\noindent\textbf{Requirements.}
In recent years, there has been a surge in research on blockchain-based control-flow automation and decision support for processes (see~\cite{Stiehle22SLR} for an overview).
\ifLong{%
The information exchanged in multi-party business processes such as the one just discussed should not be fully accessible to parties outside of those involved in the process. Unencrypted communication allows any node in the network to access the complete content of all data attached to transactions. 
If all parties shared a secret key, they could store the encrypted data using that key to ensure that only parties within their circle can read it. This would also ensure that the data is notarized by the blockchain.%
}
Typically, information shared by actors in a collaborative process is commercial-in-confidence, i.e., shared only with the parties that need access to it, and who are in turn expected to not pass the information on.
Our research complements this work by focusing on secure information exchange among multiple parties in a collaborative though partially untrusted scenario.
\ifLong{%
For instance, the purchase order of the manufacturer to the international supplier is only shared between these two actors. The customs clearance of the national customs should only be accessible to the two customs bodies (national and international) and the manufacturer.
In the case of a bidirectional communication and no need for auditability, the sender could encrypt the message with the public key of the recipient (which the sender has in a blockchain setting). 
However, this may not be practical in situations like the car production scenario where multiple parties are involved.
\todo{IW: The information regarding the previous options (no / synchronous / individual encryption) is redundant with the discussion in the introduction now.}%
}

\Cref{tab:requirements} lists the requirements stemming from the motivating use case that drives our approach and a research project in which two authors of this paper are involved.%
\footnote{Cyber 4.0 project BRIE: \url{https://brie.moveax.it/en}. Accessed: 09 June 2023.}
The table highlights the limitations of our past work~\cite{Marangone.etal/BPM2022:CAKE} that we overcome and indicates the sections in which we discuss the action taken to meet them.
Different parties should be granted access to different sections of a confidential information source (\Req{1}, as in the case of the export document in our motivating scenario).
The information source should remain available, immutable, and accountability should be granted for subsequent validations and verifications (\Req{2}, as for the check of the invoice by customs and, more in general, for process mining and auditing~\cite{Klinkmueller.etal/BCForum2019:ExtractingProcessMiningDatafromBlockchainApplications}), without major overheads (\Req{3}) for practical feasibility.
In a distributed scenario such as that of the process in \cref{sec:approach}, where multiple authorities and actors are involved, it is necessary to secure the infrastructure by avoiding that any party can acquire (\Req{4}) or forge (\Req{5}) decryption keys alone.
Finally, our approach should complement existing process execution engines to intercept and secure the data flow that characterized multi-party collaborations (\Req{6}).
Next, we discuss the background knowledge that our approach is based on.

\section{Background}
\label{sec:background}
\DLTs, and specifically programmable blockchain platforms, serve as the foundation for our work together with \MAABE. Here, we explain the basic principles 
underneath these building blocks.

\noindent\textbf{\acrlongpl{dlt}.}
\DLTs are protocols that allow for the storage, processing, and validation of transactions among a network of peers without the need for a central authority or intermediary. These transactions are timestamped and signed cryptographically, relying on asymmetric or public key cryptography with a pair of a private and a public key.
In \DLTs, every user has an account with a unique address, associated with such a key pair. 
The shared transaction list forms a ledger that is accessible to all participants in the network. 
A \textbf{blockchain} is a specific type of \DLT in which transactions are strictly ordered, grouped into blocks, and linked together to form a chain.
\DLTs, including blockchains, are resistant to tampering due to the use of cryptographic techniques such as hashing (for the backward linkage of blocks to the previous one), and the distributed validation of transactions.
These measures ensure the integrity and security of the ledger.
Blockchain platforms come endowed with consensus algorithms that allow the distributed networks to reach eventual consistency on the content of the ledger~\cite{Nakamoto/2008:Bitcoin:APeer-to-PeerElectronicCashSystem}.
Public blockchains like 
Ethereum~\cite{Wood/2018:Ethereum} 
charge fees for the inclusion and processing of transactions. 
\ifLong{%
	\ennote{Furthermore, recent blockchains implement additional features such as the \textit{multisig account}, an account managed by N multiple keys that need at least $n$, with $1 \le n \le N$, keys to sign the transaction and make it valid.} These platforms also allow the use of \textbf{Smart Contracts}, which are programs that are deployed, stored, and executed on the blockchain~\cite{Dannen/2017:IntroducingEthereumandSolidity}.%
	Ethereum and Algorand are examples of more recent blockchain protocols that support Smart Contracts.
}
Ethereum supports expressive smart contracts, which are user-defined programs.
They are deployed and invoked through transactions, i.e.,
their code is stored on chain and executed by many nodes in the network. 
Outcomes of contract invocations are part of the blockchain consensus, thus verified by the blockchain system and fully traceable.
%
The execution of smart contract code, like transactions, incurs costs measured as \emph{gas} in the Ethereum platform. Gas cost is based on the complexity of the computation and the amount of data exchanged and stored. To lower the costs of invoking smart contracts, external \PtoP systems are often utilized to store large amounts of data~\cite{Xu.etal/2019:ArchitectureforBlockchainApplications}.
One of the enabling technologies is 
\textbf{\IPFS},%
\footnote{\label{foot:ipfs} \href{https://ipfs.tech/}{ipfs.tech}. Accessed: 09 June 2023.}
a distributed system for storing and accessing files that utilizes a \DHT to scatter the stored files across multiple nodes. Like \DLTs, there is no central authority or trusted organization that retains control of all data.
\IPFS uses content-addressing to uniquely identify each file on the network. Data stored on \IPFS is linked to a resource locator through a hash, which--in a typical blockchain integration--is then sent to a smart contract to be stored permanently on the blockchain~\cite{Lopez-Pintado.etal/IS2022:ControlledFlexibilityBlockchainCollaborativeProcesses}. 
In a multi-party collaboration setting like the one presented in \cref{sec:example}, the blockchain 
provides an auditable notarization infrastructure that certifies 
transactions among the participants (e.g., purchase orders or customs clearances). Smart contracts ensure that the workflow is carried out as agreed upon, as described in~\cite{DiCiccio.etal/InfSpektrum2019:BlockchainSupportforCollaborativeBusinessProcesses,Mendling.etal/ACMTMIS2018:BlockchainsforBPM,Weber.etal/BPM2016:UntrustedBusinessProcessMonitoringandExecutionUsingBlockchain}.
Documents like purchase orders, transportation orders, and customs clearances can be stored on \IPFS and linked to transactions that report on their submission. However, data is accessible to all peers on the blockchain. To take advantage of the security and traceability of the blockchain while also controlling access to the stored information, it is necessary to encrypt the data and manage read and write permissions for specific users.

\noindent\textbf{\ABE.} \ABE is a form of public key encryption in which the \emph{ciphertext} (i.e., an encrypted version of a \emph{plaintext} message) and the corresponding decryption key 
are connected through attributes~\cite{ABE,MAABE}. In particular, \CPABE~\cite{CP-ABE,MultiAuthorityCP}
associates each potential user with a set of attributes. Policies are expressed over these attributes using propositional literals that are evaluated based on whether a user possesses a particular property.
In the following, we shall use the teletype font to format attributes and policies.
For example, user \verb|0xB0|\ldots\verb|1AA1| 
is associated with the attributes \texttt{Supplier}, to denote their role, and \texttt{43175279}, to specify the process instance number they are involved in (the \emph{case id}). For the sake of brevity, we omit from the attribute name that the former is a role and the latter a process instance identifier (e.g., \texttt{Supplier} in place of \texttt{RoleIsSupplier} or \texttt{43175279} instead of \texttt{InvolvedInCase43175279}) as we assume it is understandable from the context.
Policies are associated with ciphertexts and expressed as propositional formulae on the attributes (the literals) to determine whether a user is granted access (e.g., \texttt{Carrier or Manufacturer}).
\\
As argued in the introduction, one goal of this work is to move away from a single source of trust (or failure); thus, we consider multi-authority methods.
To decrypt and access the information in a ciphertext, a user requires a dedicated key. 
With \acrfull{maabe}, every authority creates a part of that key, henceforth \emph{\dk}. 
A \dk is a string generated via \MAABE on the basis of
\begin{iiilist}
\item the user attributes, and 
\item a secret key of the authority. 
\end{iiilist}
To generate the secret key (coupled with a public key), the authority requires public parameters composed of a sequence of pairing elements that are derived from a pairing group via Elliptic Curve Cryptography (ECC).
Due to space restrictions, we cannot delve deeper into the notions of pairing groups and pairing elements. We refer to~\cite{MAABE,miller1986use} for further details.
Once the user has obtained a \dk from every required authority, it merges them obtaining the \emph{\fdk} to decrypt the message.
\\
In the Cypertext-Policy variant of \MAABE, a ciphertext for a given message is generated from the public parameters, the public keys of all the authorities, and a policy. 
In our context, users are process participants, messages are the data artefacts exchanged during process execution, ciphertexts are encrypted versions of these artefacts, policies determine which artefacts can be accessed by which users, and keys are the tools granted to process parties to try to access the artefacts. In the following sections, we describe how we combine the use of blockchain and the Cypertext-Policy variant of \MAABE to create an access control architecture for data exchanges on the blockchain that meets the requirements listed in \cref{tab:requirements}.

\section{The MARTSIA approach}
\label{sec:approach}
In this section, we describe our approach, named Multi-Authority Approach to Transaction System for Interoperating Applications (MARTSIA).
We begin by examining the collaboration among its core software components, and then illustrate the data structures they handle.
%

\subsection{Workflow}
\label{sec:approach:workflow}
\begin{figure}[tbp]
	\includegraphics[width=.9\textwidth]{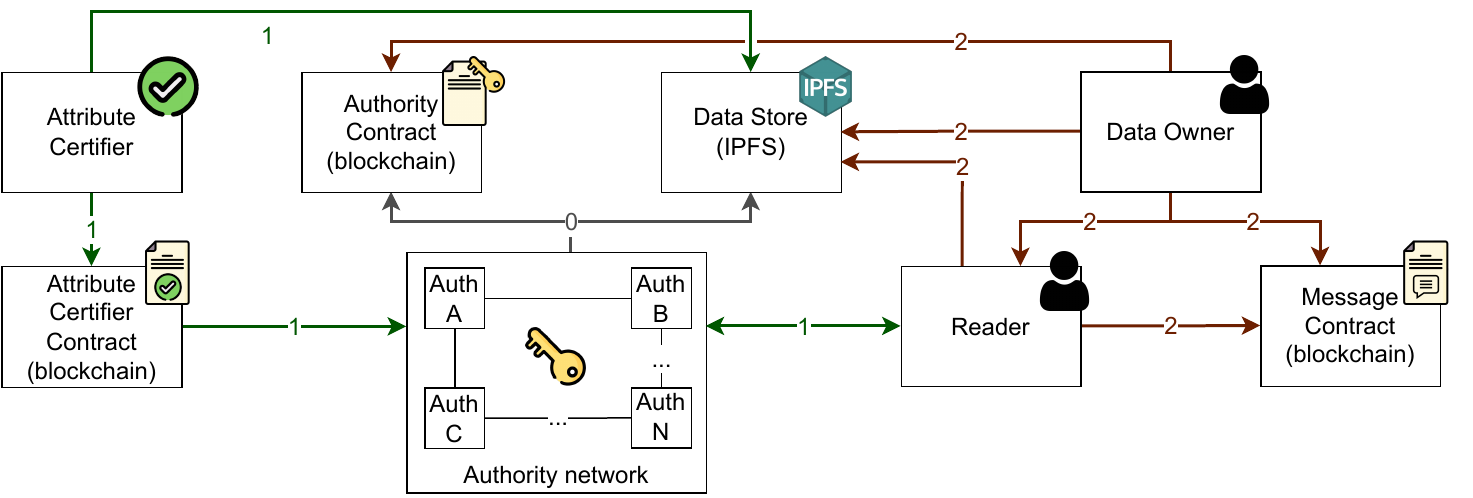}
	\caption{The key components and their interactions in the MARTSIA approach}
	\label{fig:architecture}
\end{figure}
\Cref{fig:architecture} illustrates the main components of our architecture and their interactions. 
The involved parties are:
\begin{inparadesc}
	\item[the {\AttCert}] specifying the attributes characterizing the potential readers of the information artifacts; we assume the {\AttCert}s to hold a blockchain account; different {\AttCert}s may attest to different pieces of information about potential readers;
	\item[the \DOwner] encrypting the information artifacts (henceforth also collectively referred to as \textit{plaintext}) with a specific access policy (e.g., the manufacturer who wants to restrict access to the purchase orders to the sole intended parties, i.e., the suppliers); we assume the \DOwner to hold a blockchain account and a Rivest–Shamir–Adleman (RSA)~\cite{DBLP:journals/cacm/RivestSA83} secret/public-key pair;
	\item[{\Reader}s] interested in some of the information artifacts (e.g., the manufacturer, the joystick supplier, and· the wheelchair ramp supplier); we assume the {\Reader}s to hold a blockchain account, a RSA secret-key/public-key pair, and a global identifier (GID) that uniquely identifies them; 
	\item[the \Auths] that calculate their part of the secret key for the \Reader; 
	\item[the \DStore,] a \PtoP repository based on \IPFS. \IPFS saves all exchanged pieces of information in a permanent, tamper-proof manner creating a unique content-based hash as \rloc for each of them;
	\item[the {\SC}s] used to safely store and make available the {\rloc}s to the ciphertext saved on the \DStore (\textbf{\MeC}), the information about potential readers (\textbf{\AtC}), and the data needed by the authorities to generate the public parameters (\textbf{\AutC}). \todo{ED: we are missing the SC for storing the RSA public keys}
\end{inparadesc}
%

We divide our approach in three main phases, which we discuss in detail next:
initialization (\cref{fig:workflow:authinit}), key management (\cref{fig:workflow:keymgt}), and data exchange (\cref{fig:workflow:dataex}).
In the following, the numbering scheme corresponds to the labels in \cref{fig:architecture,fig:workflow:init:keymgt,fig:workflow:dataex}.

\noindent\textbf{0: Initialization.}
Here we focus on the network of authorities, as depicted in \cref{fig:workflow:authinit}.
The initialization phase consists of the following five steps.
\begin{inparaenum}[(\bfseries{0}.1\mdseries)]
	\item \label{step:authinit:metadata} 
First, each authority creates a separate file with the metadata of all the authorities involved in the process.%
	\footnote{\label{foot:metadatalink} Notice that metadata are known to all the authorities and all the actors involved in the process.
		Therefore, non-malicious authorities are expected to create an identical file.
		The (same) hash is thus at the basis of the \rloc. 
		As a consequence, anyone can verify whether the authorities behave properly in this step by checking that the {\rloc}s are equal,
		with no need to load the file from the \DStore.}
%
Authorities are responsible for the setting of public parameters that are crucial 
to all the algorithms of \MAABE.
Therefore, we have redesigned the public parameter generation program as a \MPC protocol~\cite{Yao82b,Yao86} to guarantee full decentralization.
More specifically, we adapt a commit-then-open coin-tossing protocol~\cite{Blum81} as follows to generate a random pairing element, that is, the core piece of data described in~\cite{Rouselakis} for \MAABE implementation.
\item 
Each authority posts on the blockchain the hash of a locally generated random pairing element by invoking the \AutC.
\item 
After all the hashes are publicly stored, 
each authority posts the \emph{opening},
namely the previously hashed pairing element in-clear, completing the commit-then-open coin-tossing protocol introduced before. 
\item \label{step:init:publicparams} 
Then, every authority 
\begin{iiilist}
	\item verifies that all the hashes of the pairing elements match the respective openings, 
	\item independently combines all posted openings via bitwise XOR, and
	\item uses the output of this operation (the \emph{final shared pairing element}) to calculate the set of public parameters as illustrated in~\cite{Rouselakis}.
\end{iiilist}
\item \label{step:authinit:pkpk} 
Each authority generates its own public-key/secret-key pair by using the authority key generation algorithm of \MAABE. 
\end{inparaenum}
To enable full decentralization and notarization, we resort to the \DStore to save the output of all actions (\textbf{0.\ref{step:authinit:metadata}} to \textbf{0.\ref{step:authinit:pkpk}}) and the \AutC to keep track of the corresponding {\rloc}s.
\begin{figure}[tb]
	\centering
	\begin{subfigure}[b]{0.25\textwidth}\centering
		\includegraphics[height=0.17\textheight]{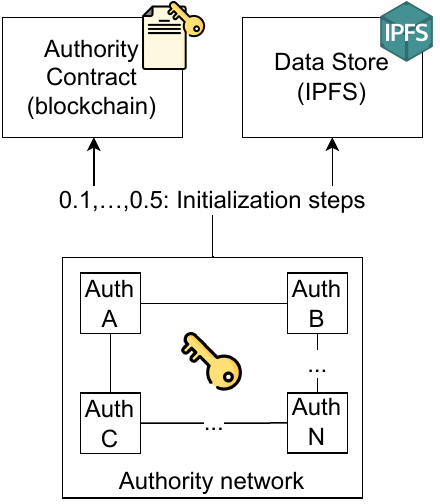}
	\caption{Authority initialization}
	\label{fig:workflow:authinit}
	\end{subfigure}
	\hfill 
	\begin{subfigure}[b]{0.7\textwidth}\centering
		\includegraphics[height=0.17\textheight]{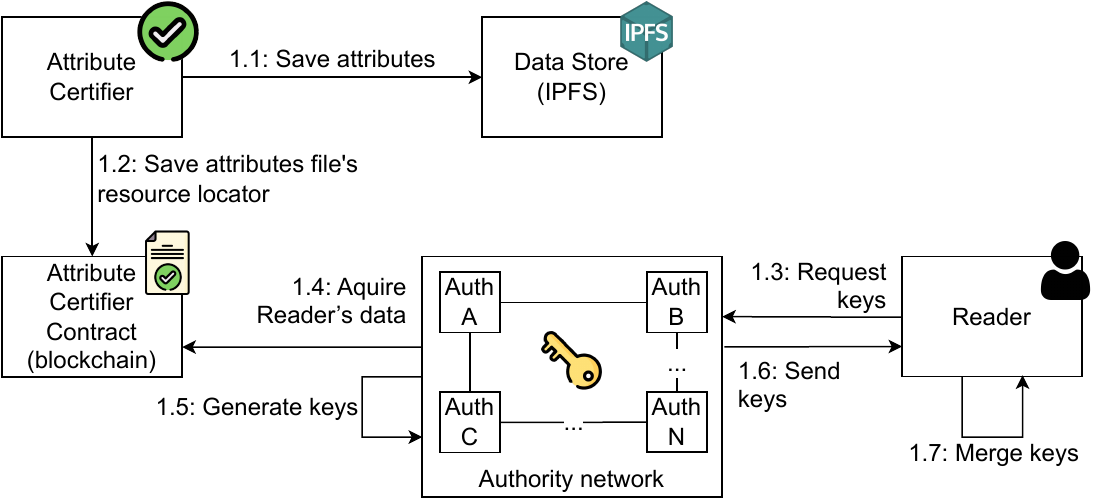}
		\caption{Key management}
		\label{fig:workflow:keymgt}
	\end{subfigure}
	\caption{Authority initialization and key management phases in MARTSIA}\label{fig:workflow:init:keymgt}
\end{figure}
\noindent\textbf{1: Key Management.}
The key management phase 
is comprised of the following steps, as illustrated in \cref{fig:workflow:keymgt}:
\begin{inparaenum}[(\bfseries{1}.1\mdseries)]
\item \label{step:key:writeattr} 
The {\AttCert}s save the attributes and the identifying blockchain account addresses of the {\Reader}s on the \DStore and
\item \label{step:key:writeattrlinkonchain} 
the corresponding \rloc on the \AtC so as to make them publicly verifiable on chain.
To this end, every \AttCert operates as a push-inbound oracle~\cite{Basile.etal/BPMBCF2021:BlockchainProcessesDecentralizedOracles}, storing on chain the attributes that determine the role of the \Reader and, optionally, the list of process instances in which they are involved.
For example, an \AttCert stores on chain that \verb|0x82|\ldots\verb|1332| is the address of a user that holds the \texttt{Manufacturer} role 
and participates in the process identified by \texttt{43175279}. 
Another \AttCert registers that \verb|0xB0|\ldots\verb|1AA1| and \verb|0x9E|\ldots\verb|C885| are {\Reader}s both endowed with the \texttt{Supplier} and \texttt{43175279} attributes, though the former is \texttt{National} and the latter \texttt{International}. 
%
\todo{Check statements\ldots}
{\Reader}s' attributes are stored on a public blockchain for verifiability.
However, notice that {\Reader}s are referred to by their public addresses, thus keeping pseudonimity. Also, the attribute names are strings that we keep intuitively understandable for the sake of readability in this paper, yet only serve as propositional symbols for the encoding of policies. Obfuscation techniques can thus be seamlessly applied though their discussion goes beyond the scope of this paper.
\todo{\ldots up to here.}
Whenever a {\Reader} (e.g., the international customs) wants to access the data of a message (e.g., the sections of interest in the Document), they operate as follows:
\item \label{step:key:request} 
They request a key to all the authorities, passing the identifying GID as input (we enter the detail of this passage later);
\item 
Each authority seeks the \Reader data (the blockchain address and attributes), and 
obtains them from the \AtC;
\item 
Equipped with these pieces of information alongside the public parameters, the secret key, and the user's GID, each authority produces a \MAABE decryption key (\dk) 
for the \Reader, and 
\item 
sends it back. Once all {\dk}s are gathered from the authorities, 
\item 
the \Reader can merge them to assemble their own \fdk. 
Notice that none of the \Auths can create the \fdk alone (unless specified as such), thus meeting \Req{5}; no user other than the intended \Reader can obtain the key (\Req{4}).
The key management phase can be interleaved with the Data Exchange phase (below).
\end{inparaenum}

\noindent\textit{A note on the security of key requests.} 
Maliciously obtaining the \fdk of another \Reader is a high security threat. We decompose this issue in two challenges. First, we want to avoid that information exchanged between a \Reader and an \Auth is intercepted by other parties. To this end, we convey every communication in step~\textbf{1.\ref{step:key:request}} through a separate client-server \gls{ssl} connection between the \Reader and each of the authorities.
To avoid any false self-identification as a \Reader through their GID, we include a handshake preliminary phase in the protocol.
It starts with every \Auth (server) sending a random value (or \textit{challenge}) to the \Reader (client). The latter responds with that value signed with their own RSA private key, so as to let the invoked components verify their identity with the caller's public key.
\ifLong{%
	\noindent{\textbf{A note on key requests.}}
	\todo{Come far too late. To be probably divided in two parts. The first one crunched in a few lines and moved to where this phase really happens (1). The second one spent as future work.}
	%
	The \Reader can produce a key request to the authorities in two ways.
	The first one is an client-server connection between the \Reader and all the authorities, one by one.
	Once all the \Auth servers are invoked and the \Reader has obtained all the parts of the secret key, they merge the {\dk}s to obtain the \fdk and decrypt the symmetric key.
	We remark that the communication backbone outside of the blockchain and the \DStore for the information exchanges between components is based on the \gls{ssl} protocol, so as to avoid packet sniffing from malicious third parties that could intercept the data.
	To avoid that any malicious peer could request the \dk in place of the real \Reader by knowing their address, we designed a preliminary handshake phase, which starts with every \Auth (server) sending a random value to the \Reader (client). The latter responds with that value signed with their own private key, so as to let the invoked components verify their identity with the caller's public key.
	The second way for a \Reader to make a key request to the authorities is through the blockchain. In this case, the \Reader makes a key request sending a transaction directly to the intended authority. In the data payload of the transactions, the \Reader specifies the necessary metadata
	\todo{What are these necessary metadata?}
	for the authority and then awaits for a response transaction from the invoked authority. This authority sees the transaction with the key request on the blockchain and starts the key generation phase. After some
	\todo{Some? A dozen? Random?}
	security checks, it generates the secret key.
	\todo{Maybe the \dk?}
	Then, the \Auth encrypts it with the RSA public key of the user and stores the result in a \DStore file. Finally, the \Auth sends the corresponding hash (the \rloc) through the blockchain to the \Reader that has invoked it via transaction. The \Reader gets the \rloc from the data payload and 
	retrieves the file from the \DStore. Finally, they decrypt the content with their RSA private key. 
	Once they have obtained all {\dk}s, they can merge them to forge the {\fdk}.
}%

\begin{figure}[tb]
	\centering
	\includegraphics[height=0.14\textheight]{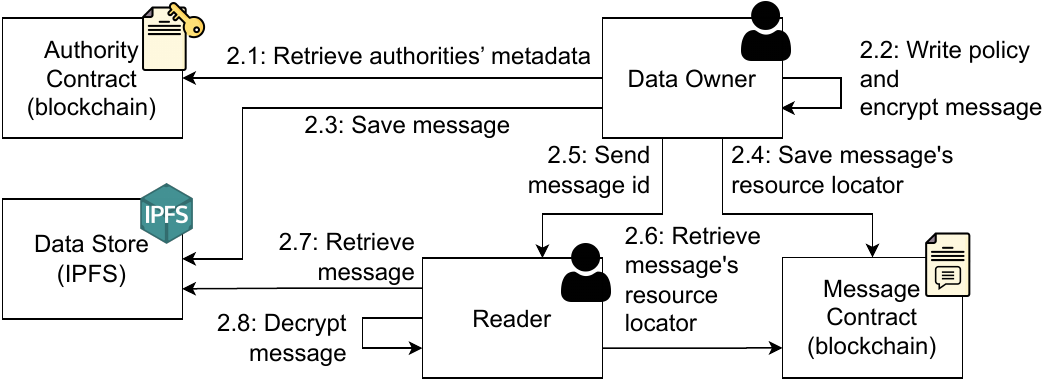}
	\caption{The data exchange phase in MARTSIA}
	\label{fig:workflow:dataex}
\end{figure}
%
\noindent\textbf{2: Data Exchange.}
\Cref{fig:workflow:dataex} presents the operations carried out for information storage and access.
As a preliminary operation, the \DOwner verifies that the hash links of the files with metadata and public parameters posted by all the authorities are equal to one another to ascertain their authenticity.\footref{foot:metadatalink}
Then, data is transferred from the \DOwner to the {\Reader}s through the following steps.
\begin{inparaenum}[(\bfseries{2}.1\mdseries)]
\item 
\label{step:dataex:encrypt}  
The \DOwner retrieves the authorities' public keys and public parameters from the blockchain.
\item \label{step:dataex:policyenc} 
Then, they write a policy.
Notice that standard \MAABE sets a maximum length for the input files.
In a business process context, this limitation would undermine practical adoption.
To cater for the encryption of arbitrary-size plaintexts, we thus resort to a two-staged hybrid encryption strategy~\cite{CramerS03}.
First, the \DOwner encrypts via \MAABE a randomly generated symmetric key (of limited size, e.g., \verb|b'3:go+s|\ldots\verb|x2g='|) with the authorities' public keys and the policy (obtaining, e.g., \verb|eJytm1|\ldots\verb|eaXV2u|). 
Afterwards, it encrypts the actual information artifact (of any size) via symmetric key encryption scheme~\cite{NIST/FIPS197-2001:AES} using that symmetric key.
In our example scenario, then, the manufacturer does not encrypt via \MAABE the supply purchase order, but the key through which that document is encrypted (and decryptable). 
\item 
Thereupon, the \DOwner (e.g., the manufacturer) saves the encrypted symmetric key and information artifact (plus additional metadata we omit here for the sake of clarity and detail in \cref{sec:approach:datastructures}) in one file on the \DStore, 
\item \label{step:dataex:policyenconchain} 
sends the file's \rloc to the \MeC, and
\item 
transmits the unique message ID (e.g., \texttt{22063028}) assigned to the file to the \Reader (e.g., the supplier).
As the information artifact is on the \DStore and its \rloc saved on chain, it is written in a permanent, tamper-proof and non-repudiable way, thus meeting requirement \Req{2}.
%
Equipped with their own \fdk,
the \Reader can begin the message decryption procedure.
\item 
At first, the \Reader retrieves the \rloc of the message from the \MeC. 
\item 
Then, once the \Reader obtains the ciphertext from the \DStore, they pass it as input alongside the public parameters (see step \textbf{0.\ref{step:init:publicparams}} above) and the \fdk to the \MAABE decryption algorithm running locally. 
\item 
Mirroring the operations explained in step \textbf{2.\ref{step:dataex:encrypt}}, \MAABE decrypts the symmetric key from the retrieved ciphertext. Only with the symmetric key, the \Reader can obtain the original information artifact. 
\end{inparaenum}
\subsection{Data Structures}
\label{sec:approach:datastructures}
After the analysis of the software components and tasks employed in our approach, we focus on its core data structures: messages and policies.

\begin{table}[tbp]
	\caption[Example of messages stored by MARTSIA upon encryption]{Example of messages stored by MARTSIA upon encryption \vspace{-0.75em}}
	\label{tab:messageEncoding:after}
	\resizebox{\columnwidth}{!}{%
		\begin{tabular}{|c|l|l|l|}
\hline
   Message
 & Metadata
 & Header
 & Body (slices) \\ \hline
 
   \makecell{Supply \\ purchase \\ order \\ (ramp)}
 & 
\begin{lstlisting}[firstline=1,lastline=3]
sender:              0x82[...]1332
process_instance_id: 43175279
message_id:          22063028
\end{lstlisting}
 & 
\begin{lstlisting}[firstline=2,lastline=4]
EncryptedKey: eJytm1[...]eaXV2u
Fields:      {CV8=[...]6w==, p84W[...]aw==,
              avNL[...]lw==, AOo=[...]mQ==}
\end{lstlisting}
 & 
\begin{lstlisting}[firstline=2,lastline=5]
CV8=[...]6w==:  Ruk/[...]QQ==
p84W[...]aw==:  1UTk[...]mA==
avNL[...]lw==:  VT+D[...]JQ==
AOo=[...]mQ==:  ZUE/[...]6w==
\end{lstlisting} \\ 
 \hline
  
 & 
 & 
\begin{lstlisting}[firstline=8,lastline=14]
SliceId:      62618638
EncryptedKey: eJytWU[...]Bmn4k=
Fields:      {u7o=[...]iw==, TacL[...]IQ==, 
              EhrB[...]lw==, Tiba[...]Wg==, 
              xJ+6[...]SQ==, RPn7[...]Zz==, 
              T3Wq[...]eK==, QXjN[...]LS==, 
              Jmk8[...]fL==}
\end{lstlisting}
 & 
\begin{lstlisting}[firstline=8,lastline=16]
u7o=[...]iw==:  py8l[...]mA==
TacL[...]IQ==:  MKNH[...]uQ==
EhrB[...]lw==:  5bz0[...]bg==
Tiba[...]Wg==:  JahD[...]5Q==
xJ+6[...]SQ==:  3aIm[...]kg==
RPn7[...]Zz==:  nRNh[...]GA==
T3Wq[...]eK==:  PxRk[...]0g==
QXjN[...]LS==:  Lfhk[...]Ug==
Jmk8[...]fL==:  Gj43[...]yg==
\end{lstlisting} \\ 
 \cline{3-3}
 \makecell{Export \\ document}
 & 
\begin{lstlisting}[firstline=5,lastline=7]
sender:              0x9E[...]C885
process_instance_id: 43175279
message_id:          15469010
\end{lstlisting}
 & 
\begin{lstlisting}[firstline=16,lastline=20]
SliceId:      19756540
EncryptedKey: eJytm0[...]wtYFo=
Fields:      {ZOSH[...]9B==, BZ8l[...]KY==,
              QeUy[...]WT==, rtdZ[...]hf==,
              8kuG[...]Ts==}
\end{lstlisting}
 & 
\begin{lstlisting}[firstline=17,lastline=21]
ZOSH[...]9B==:  sctC[...]nQ==
BZ8l[...]KY==:  5Gln[...]Pg==
QeUy[...]WT==:  k49+[...]kA==
rtdZ[...]hf==:  lRLd[...]tm==
8kuG[...]Ts==:  25E0[...]uc==
\end{lstlisting} \\ 
 \cline{3-3}
   
 & 
 & 
\begin{lstlisting}[firstline=22,lastline=25]
SliceId:      12191034
EncryptedKey: eJytWU[...]sGlU4=
Fields:      {t8gr[...]QQ==, yuwd[...]vg==,
              1K1d[...]zQ==}
\end{lstlisting} 
 & 
\begin{lstlisting}[firstline=22,lastline=24]
t8gr[...]QQ==:  ZJ1v[...]5A==
yuwd[...]vg==:  7XpN[...]4A==
1K1d[...]zQ==:  +QbM[...]Cw==
\end{lstlisting} \\ 
 \cline{3-3}
 
 &  
 & 
\begin{lstlisting}[firstline=27,lastline=31]
SliceId:      98546521
EncryptedKey: eJytk9[...]oIJS6=
Fields:      {rjY=[...]KQ==, ZdWC[...]xg==,
              6aLB[...]iw==, VD2h[...]6w==,
              8UmX[...]MQ==}
\end{lstlisting}
 & 
\begin{lstlisting}[firstline=25,lastline=29]
rjY=[...]KQ==:  JPAv[...]LA==
ZdWC[...]xg==:  w05J[...]Hg==
6aLB[...]iw==:  0wWu[...]vA==
VD2h[...]6w==:  eZu7[...]QQ==
8UmX[...]MQ==:  sXaB[...]kQ==
\end{lstlisting} \\
 \hline
 
 \makecell{National \\ customs \\ clearance}
 & 
\begin{lstlisting}[firstline=9,lastline=11]
sender:              0x5F[...]FFE1
process_instance_id: 43175279
message_id:          64083548
\end{lstlisting}
 & 
\begin{lstlisting}[firstline=35,lastline=38]
EncryptedKey: eJytWU[...]gGl2Y=
Fields:      {fdoT[...]kA==, 2AkH[...]Rw==,
              0bTn[...]6A==, RZVJ[...]rQ==,
              4TXI[...]zw==}
\end{lstlisting}
 & 
\begin{lstlisting}[firstline=32,lastline=36]
fdoT[...]kA==:  fUSZ[...]Bg==
2AkH[...]Rw==:  dGp2[...]zA==
0bTn[...]6A==:  TuR9[...]bA==
RZVJ[...]rQ==:  Pq8U[...]dQ==
4TXI[...]zw==:  OIzx[...]Mw==
\end{lstlisting} \\
 \hline
 
\end{tabular}

	}
\end{table}

\begin{table}[tbp]
	\caption[Message policy examples]{Message policy examples \vspace{-0.75em}}
	\label{tab:messagePolicies}
	\resizebox{1\columnwidth}{!}{%
		\begin{tabular}{|c|c|c|}
\hline
Message & Slice &  Policy \\ \hline
 \makecell[c]{Supply purchase \\ order (ramp)}
 &   & 
\begin{lstlisting}[firstline=1,lastline=1]
43175279@2+ and (Manufacturer@1+ or (Supplier@1+ and International@1+))
\end{lstlisting}\\ 
\hline
 & \ref{item:intshipmentorder} &
\begin{lstlisting}[firstline=2,lastline=3,linewidth=1.5\columnwidth]
Customs@A or (43175279@2+ and ((Supplier@1+ and International@1+) or 
                                 Manufacturer@1+ or (Carrier@1+ and International@1+)))
\end{lstlisting}\\ \cline{2-3}
 Export & \ref{item:csdd} &
\begin{lstlisting}[firstline=4,lastline=4]
Customs@A or (43175279@2+ and (Supplier@1+ and International@1+))
\end{lstlisting}\\ \cline{2-3}
 document & \ref{item:orderreference} &
\begin{lstlisting}[firstline=5,lastline=5]
43175279@2+ and ((Supplier@2+ and International@1+) or Manufacturer@1+)
\end{lstlisting}\\ \cline{2-3} 
 & \ref{item:invoice} &
\begin{lstlisting}[firstline=6,lastline=6,linewidth=1.5\columnwidth]
Customs@A or (43175279@2+ and ((Supplier@1+ and International@1+) or Manufacturer@1+))
\end{lstlisting} \\
 \hline
 \makecell[c]{National customs \\ clearance}
 &   & 
\begin{lstlisting}[firstline=7,lastline=7]
Customs@A or (43175279@2+ and Manufacturer@1+)
\end{lstlisting} \\ \hline
\end{tabular}
	}
\end{table} 

\noindent{\textbf{Messages.}}
\Cref{tab:messageEncoding:before} illustrates the messages we described in our running example in~\cref{sec:example} along with a generated symmetric key for each message. 
\Cref{tab:messageEncoding:after} shows the messages as saved on the \DStore by the \DOwner after the encryption process explained in \cref{sec:approach:workflow} (phase 2).
Each file stored on the \DStore consists of one or more sections to be accessed by different actors (henceforth, \emph{slices}). Every slice is divided in three parts.
\begin{inparadesc}
	\item[The metadata] contain the \emph{message sender} (e.g., \verb|0x82|\ldots\verb|1332| in \cref{tab:messageEncoding:after}), the \emph{case id} (e.g., \texttt{43175279}), and the \emph{message id} that uniquely identifies the message (e.g., \texttt{22063028}).
	\item[The body] is the encrypted information saved as key/value entries (\emph{fields}) for ease of indexation. For security, notice that neither the keys nor the values are in clear. 
	\item[The header] consists of the \emph{encrypted} symmetric \emph{key} generated at step~\textbf{2.\ref{step:dataex:encrypt}}, and the list of field keys that the body contains. In case two or more slices form the message (as in the case of the export document), each is marked with a unique \emph{slice id} (e.g., \texttt{62618638}).  
\end{inparadesc}
We recall that a message is stored on the \DStore and retrievable through a hash, content-based \rloc. The \rloc can thus be attached to process execution data for monitoring and auditability purposes, in compliance with \Req{6}.

\begin{sloppypar}
\noindent{\textbf{Policies.}} We use \MAABE policies to specify read grants to message slices, thus enabling fine-grained access control as per \Req{1}. For example, the export document written by the international supplier of process instance \texttt{43175279} is partitioned in four slices as illustrated in \cref{tab:messageEncoding:before}. \Cref{tab:messagePolicies} shows the encoding of the policies that restrict access to specific classes of {\Reader}s, based on the attributes the {\AttCert}s attested to in step~\textbf{1.\ref{step:key:writeattr}}. 

Henceforth, we will use the following notation to encode a policy $P$.
We shall use $\mathrm{Attr}$\texttt{@}$X$ as a shorthand notation for a policy indicating that an authority $\mathrm{Auth}$ (if $X$ is $\mathrm{Auth}$) or \emph{at least} $n \geq 1$ authorities (if $X$ is $n$\texttt{+}) 
generate the key based on the verification of attribute $\mathrm{Attr}$. 
Compound policies can be formed by joining $\mathrm{Attr}$\texttt{@}$X$ propositions with \texttt{or} and \texttt{and} logical operators.
For instance, \texttt{(Customs@A or Supplier@1+)} declares that only authority \texttt{A} 
can authorize customs, whereas \emph{any} authority can generate the \dk for suppliers,
and that only customs or suppliers can read a message. 
To sum up, we will henceforth use the following grammar for policies $P$:\vspace{-5pt}
\begin{align*}
	P ::=\: & \mathrm{Attr}\texttt{@}X && \mid \quad  P \texttt{ and } P' &&  \mid \quad  P \texttt{ or } P' \quad \textit{ where }\mathrm{Attr}\textit{ is an attribute;}\\
	X ::=\: & \mathrm{Auth}            && \mid \quad  n\texttt{+}         && \textit{ where }\mathrm{Auth}\textit{ is an authority and }n\textit{ a positive integer.}
\end{align*}

\vspace{-5pt}
Notice that we enable the selection of a specific \dk forger for backward compatibility towards single-authority frameworks.
The downsides are that
\begin{iiilist}
	\item no key is generated if that authority is down (if \texttt{A} crashed, e.g., a user cannot be recognized as a customs body), and
	\item a corrupted authority could take over the generation of an {\fdk} if only one attestation is necessary (theirs).
\end{iiilist}
Therefore, special attention must be paid in the writing of policies.
\texttt{43175279@2+} requires that \emph{at least} two authorities attest to the participation of a user in case \texttt{43175279}. 
%
A user that is not authorized by all the required authorities cannot have 
the \acrlong{fdk} as per the policy. Also, 
whenever multiple authorities are involved in the generation of the \fdk by contributing to a part of it (the \dk),
only the user can compose the \fdk
and decrypt the ciphertext.
\todo{
	IW: Cool, thanks for fixinng! Just one question now: how / where to specify who's an authority? (Otherwise I'll create 5 authorities before grabbing lunch now ;-)
}

In our example, the international shipment order is the first slice of the export document. It should be readable by the national and international customs, and by specific actors involved in the process instance: the sender (i.e., the international supplier), the manufacturer, and the international carrier. Additionally, we exert constraints on the authorities providing the \dk: Customs 
are given the \dk by \Auth~\texttt{A}, 
and at least two \Auths must declare that a \Reader is involved in the given process instance.
The other attributes can be attested to by any \Auth.
This composite rule translates to the following expression:
\texttt{Customs@3+ or (43175279@2+ and ((Supplier@1+ and International@1+) or}
\texttt{Manufacturer@1+ or (Carrier@1+ and International@1+)))}.
\end{sloppypar}
\todo{DONE: Added 'flow' between sections 4 and 5}

Thus far we have described the architecture of MARTSIA, along with its operations, employed techniques and data structures. Next, we focus on its realization and testing.

\section{Implementation and evaluation}
\label{sec:imptes}
MARTSIA is an approach aimed at securing the access to information at a fine-grained level in a distributed fashion.
We have hitherto shown its security guarantees by design, using a multi-party process execution as a motivating scenario.
In this section, we experimentally evaluate whether MARTSIA can deliver its guarantees and properties in a process context, and at what cost.
%
The code of our prototype alongside the detailed results of our experiments can be found at
\href{https://github.com/apwbs/MARTSIA-Ethereum}{github.com/apwbs/MARTSIA-Ethereum}.
We implemented the three contracts described in \cref{sec:approach:workflow} as a single instance in Solidity, a programming language for the Ethereum Virtual Machine (EVM).
We deployed the instance on the Sepolia (Ethereum), Mumbai (Polygon), and Fuji (Avalanche) testnets.%
\footnote{Goerli: \href{https://sepolia.etherscan.io/}{sepolia.etherscan.io}; Mumbai: \href{https://mumbai.polygonscan.com/}{mumbai.polygonscan.com}; Avalanche: \href{https://testnet.snowtrace.io/}{testnet.snowtrace.io}. Accessed: 09 June 2023.}
We created an \IPFS local node%
\footref{foot:ipfs}
to realize the \DStore, and used Python to encode the off-chain modules including the client-server communication channels.
%

\begin{figure}[tbp]
	\includegraphics[width=\textwidth]{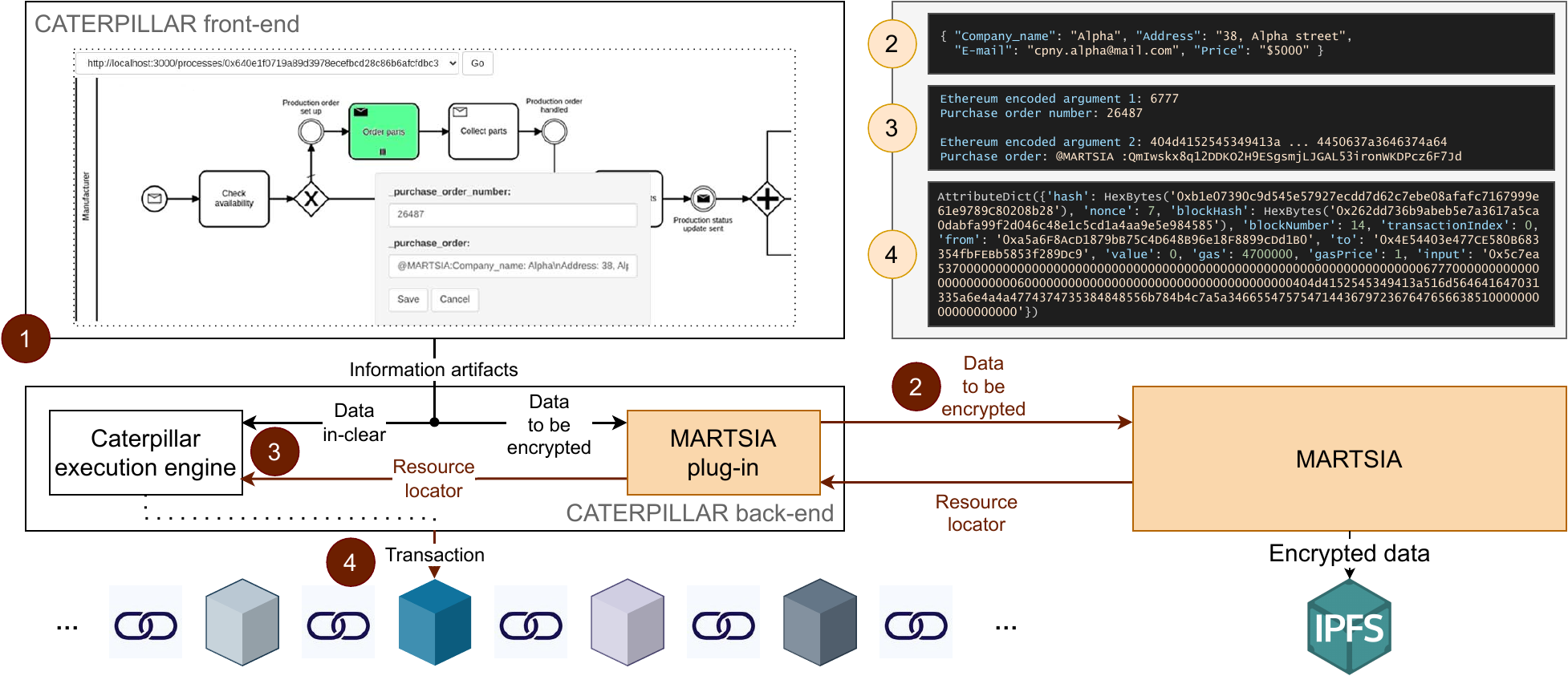}
	\caption{A run of our integration of MARTSIA with Caterpillar~\cite{Lopez-Pintado.etal/SPE2019:Caterpillar}}
	\label{fig:martsia-x-caterpillar}
\end{figure}

First, to demonstrate the adoption of MARTSIA as a secure data-flow management layer for process execution (thus meeting requirement \Req{6}), we created an integration module connecting our tool with a state-of-the-art blockchain-based process execution engine tool, i.e., Caterpillar~\cite{Lopez-Pintado.etal/SPE2019:Caterpillar}. 
For our experiments, we used Caterpillar v.1.0.%
\footnote{\href{https://github.com/orlenyslp/Caterpillar}{github.com/orlenyslp/Caterpillar}. Accessed: 10 June 2023.}
The code to replicate and extend our test is publicly available in our repository.%
\footnote{\href{https://github.com/apwbs/MARTSIA-Ethereum/tree/main/caterpillar-interaction}{github.com/apwbs/MARTSIA-Ethereum/tree/main/caterpillar-interaction}}
As shown in \cref{fig:martsia-x-caterpillar}, we insert a plug-in in the architecture of Caterpillar to use MARTSIA as an intermediate data layer manager, replacing the built-in data store for information securing.
We illustrate our experiment with a simplified fragment of the running example (see \cref{sec:example}) focusing on the purchase order sent from the manufacturer to the international supplier. The exchanged data artifact consists of a purchase order number, to be publicly stored on chain, and the confidential purchase order entries listed in the top row of \cref{tab:messageEncoding:before}. The user passes both the entries as input through the Caterpillar panel (see mark~{\CircOne} in \cref{fig:martsia-x-caterpillar}), specifying the data they want to be secured by MARTSIA with a special prefix (``\texttt{@MARTSIA:}'').
\todo{Recheck numbering as now boxes represent a different thing.}
Our integration module captures the input and encrypts the indicated entry as explained in \cref{sec:approach} so that only the supplier can read and interpret those pieces of information as per \Req{1} (\CircTwo). Once the encryption is concluded, MARTSIA invokes the Caterpillar Smart Contract passing the first argument (the purchase order number) as is, and replaces the second argument with a marked \IPFS link in place of the original data (\CircThree). The resource locator for the stored information is thus saved on the ledger by the process execution engine for future audits (\CircFour), yet not publicly readable (\Req{2}). Thereupon, the recipient of the confidential information (or the auditor, later on) can retrieve and decode the information with their secret key (\Req{4}) provided by the authority network (\Req{5}). 

Aside from empirically showing the suitability of MARTSIA as a secure data-flow manager in an ensemble with a process execution engine, we remark that the cost overhead in terms of transaction fees required by MARTSIA is negligible with respect to the main process execution management. For example, the method running the activity \texttt{Order parts} of the BPMN in~\cref{fig:example}
on Caterpillar incurs \num{114494} gas units for execution with our inputs.
The on-chain component of MARTSIA, detached from Caterpillar, requires \num{89772} gas units to store the \IPFS link. As we use the second string input field of the Caterpillar's smart contract to save that resource locator, the separate gas consumption for MARTSIA is unnecessary and can be directly included in the overall process execution costs. Notice that the same activity execution saving the purchase order as plaintext (thus renouncing to the fine-grained confidentiality guarantees of MARTSIA) would 
have entailed a larger cost because the textual content is a longer string than an \IPFS link: \num{116798} gas units.
Auditability on process execution \emph{and} secure message exchanges are thus guaranteed with low overhead, as stated in \Req{3}.
\begin{table}[tb]
\caption[Execution cost and timing of the steps that require an interaction with the blockchain]{Execution cost and timing of the steps that require an interaction with the blockchain \vspace{-0.75em}}%
\label{tab:exec:cost}%
\resizebox{1.0\columnwidth}{!}{%
    \begin{tabular}{l
  S[table-format = 6.3,round-precision = 3,round-mode=places]
  S[table-format = 7.3,round-precision = 3,round-mode=places]
  S[table-format = 5.3,round-precision = 3,round-mode=places]
  S[table-format = 6.3,round-precision = 3,round-mode=places]
  S[table-format = 4.3,round-precision = 3,round-mode=places]
}
    \toprule
      {~} &
      \multicolumn{4}{c}{\textbf{Execution cost} $\left[\mathrm{Gwei}=\mathrm{ETH} \times10^{-9}\right]$} &
      \\
      \cmidrule{2-5}
      {~} &
      \multicolumn{1}{r}{{Contract deployment}} & 
      \multicolumn{1}{c}{{Steps~\textbf{0.\ref{step:authinit:metadata}}~to~\textbf{0.\ref{step:authinit:pkpk}}}} & 
      \multicolumn{1}{c}{{Step~\textbf{1.\ref{step:key:writeattrlinkonchain}}}} & 
      \multicolumn{1}{c}{{Step~\textbf{2.\ref{step:dataex:policyenconchain}}}} &
      \\ 
      \multicolumn{1}{l}{\textbf{Platform}} &
      \multicolumn{1}{c}{(\num{1692955} gas units)} &
      \multicolumn{1}{c}{(\num{476547} gas units)} &
      \multicolumn{1}{c}{(\num{67533} gas units)} &
      \multicolumn{1}{c}{(\num{89772} gas units)} &
      \textbf{Avg.\ latency}~{[\si{\milli\second}]}\\
      \cmidrule(r){1-1}
      \cmidrule{2-5}
      \cmidrule(l){6-6}
      \textbf{Sepolia} \textrm{(ETH)} &
       2539432.51354364e-09 &
       714820.503880085e-09 &
       101299.500540264e-09 &
       134658.000718176e-09 &
       9288.573646942760 \\
      \textbf{Fuji} \textrm{(AVAX)} &
       340498.77068816e-09 &
       95873.484581149e-09 &
       13586.538284745e-09 &
       18060.662362105e-09 &
       4278.099486033120 \\
      \textbf{Mumbai} \textrm{(MATIC)} &
       1283.162694703e-09 &
       354.691115093e-09 &
       50.311010581e-09 &
       66.011515441e-09 &
       4944.807272752120 \\
       \midrule
       &
       \multicolumn{4}{c}{\textbf{Off-chain execution time}~{[\si{\milli\second}]}}
       & \\
       \cmidrule{2-5}
       &
       0.000 &
       2582.47053623199 &
       38.27977180481 &
       158.4467888 &
       \\
    \bottomrule
\end{tabular}
}%
\end{table}

To gauge the gas expenditure and execution time of our system's on-chain components, we called the methods of the deployed {\SC} daily on Sepolia, Fuji and Mumbai for \num{14} days (from 16 to 29 May 2023).
All the experiment's transactions and gas measurements are available in our code repository.%
\footnote{\href{https://github.com/apwbs/MARTSIA-Ethereum/tree/main/tests}{github.com/apwbs/MARTSIA-Ethereum/tree/main/tests}. Accessed: 10 June 2023.}
The data we used to run the tests are taken from our running example (see \cref{sec:example}).
\Cref{tab:exec:cost} illustrates the results.
We divide the measurement in four different phases: 
\begin{iiilist}
	\item the deployment of the \SC;
	\item the initialization of the {\Auths} (steps~\textbf{0.\ref{step:authinit:metadata}}~to~\textbf{0.\ref{step:authinit:pkpk}});
	\item the {\Reader} certification (step~\textbf{1.\ref{step:key:writeattrlinkonchain}});
	\item the storage of a message by a \DOwner to save a message (step~\textbf{2.\ref{step:dataex:policyenconchain}}).
\end{iiilist}
For all the above phases, the table shows the average gas consumed for execution (ranging from \num{67533} units for step~\textbf{1.\ref{step:key:writeattrlinkonchain}} to \num{1692955} units for the \SC deployment) and the cost converted in Gwei (i.e., $10^{-9}$ Ether).%
\todo{EDO+CDC:Time cost analysis added}
Along with the analysis of costs, we 
measured the time needed to perform the steps of our approach. Each step involves sending a transaction to the blockchain. Therefore, we separate the execution time between the off-chain data elaboration and the latency induced by the blockchain infrastructure. The average time required to store a transaction in a block ranges from approximately \SI{4.3}{\sec} (Fuji) to \SI{9.3}{\sec} (Sepolia). 
The off-chain passages require a lower time, from about \SI{0.038}{\sec} for the {\Reader} certification ({Step~\textbf{1.\ref{step:key:writeattrlinkonchain}}}) to the circa \SI{2.6}{\sec} needed for the cooperative work carried out by the {\Auths} during the initilization phase (steps~\textbf{0.\ref{step:authinit:metadata}}~to~\textbf{0.\ref{step:authinit:pkpk}}).
More in-depth comparative analyses and a stress test of the architecture 
pave the path for future endeavors, as we discuss in \cref{sec:conclusion} after a summary of the state of the art.

\section{Related work}
\label{sec:sota}
\ifShort{%
In recent years, numerous approaches have been proposed to automate collaborative processes using blockchain technology~\cite{DiCiccio.etal/InfSpektrum2019:BlockchainSupportforCollaborativeBusinessProcesses} beyond the aforementioned Caterpillar~\cite{Lopez-Pintado.etal/SPE2019:Caterpillar}.
Previous studies in the area have shown the effectiveness of blockchain-based solutions to add a layer of trust among actors in multi-party collaborations~\cite{Weber.etal/BPM2016:UntrustedBusinessProcessMonitoringandExecutionUsingBlockchain} even in adversarial settings~\cite{Madsen.etal/FAB2018:CollaborationamongAdversaries:DistributedWorkflowExecutiononaBlockchain}, improve verifiability of workflows with model-driven approaches~\cite{Lopez-Pintado.etal/SPE2019:Caterpillar,Tran.etal/BPMDemos2018:Lorikeet}, allow for monitoring~\cite{DiCiccio.etal/SoSyM2022:BlockchainForProcessMonitoring}, mining~\cite{Klinkmueller.etal/BCForum2019:ExtractingProcessMiningDatafromBlockchainApplications}, and auditing~\cite{Corradini.etal/ACMTMIS2022:EngineeringChoreographyBlockchain}.
Interestingly, a more recent release of Caterpillar~\cite{Lopez-Pintado.etal/IS2022:ControlledFlexibilityBlockchainCollaborativeProcesses} enables the dynamic allocation of actors based on a language for policy bindings. 
MARTSIA has the capability to adjust roles dynamically as well, as access keys are created based on actors' attributes verified at runtime.
}%
\ifLong{%
In recent years, efforts have been made to automate collaborative processes using blockchain technology.
Weber et al.~\cite{Weber.etal/BPM2016:UntrustedBusinessProcessMonitoringandExecutionUsingBlockchain} first developed a method that uses blockchain to facilitate business processes between untrusted parties. Their work demonstrates that actors can reach a mutual agreement without relying on a central authority for enforcement, thanks to the use of blockchain.
Tran et al.~\cite{Tran.etal/BPMDemos2018:Lorikeet} developed Lorikeet, a model-driven engineering tool that implements business processes on the blockchain for asset management, providing a solution for scenarios that usually require central authorities.
Di Ciccio et al.~\cite{DiCiccio.etal/InfSpektrum2019:BlockchainSupportforCollaborativeBusinessProcesses} outline the design and execution of business processes involving multiple parties. They also discuss the key components of model-driven approaches for blockchain-based collaborative business processes and compare Caterpillar and Lorikeet.
López Pintado et al. in~\cite{Lopez-Pintado.etal/IS2022:ControlledFlexibilityBlockchainCollaborativeProcesses} propose a model for dynamically allocating actors in a multi-party business process to roles and creating a language for policy bindings.
MARTSIA has the capability to adjust roles dynamically, with attributes being determined by the Attribute Certifier either during runtime or deployment. Access keys are only created when requested, not before the process begins.

Madsen et al.~\cite{Madsen.etal/FAB2018:CollaborationamongAdversaries:DistributedWorkflowExecutiononaBlockchain} examine distributed workflow execution that involves collaboration between adversaries. In such scenarios, the participating parties do not trust one another and may also suspect that someone may not follow the established protocols.
The authors show that a Smart Contract can be used to implement the execution of a distributed declarative workflow, while still maintaining enforcement of its semantics and recording the execution history.
Corradini et al.~\cite{Corradini.etal/ACMTMIS2022:EngineeringChoreographyBlockchain} introduce ChorChain, a tool that transforms a BPMN choreography model into a Solidity Smart Contract. ChorChain also provides auditors with access to information on process instances, both after and during runtime.%
}
These studies enhance the integration of blockchain technology with process management, unlocking security and traceability benefits. However, they primarily focus on the control-flow perspective and lack mechanisms for secure access control to data stored on public platforms. In contrast, our work focuses specifically on this aspect in the context of collaborative business processes and, as we demonstrated in \cref{sec:imptes}, can complement existing blockchain-based process execution engines.

Another area of research related to our investigation is the protection of privacy and integrity of data stored on the blockchain. Several papers in the literature explore the use of encryption for this purpose. Next, we provide an overview of techniques. Hawk~\cite{Hawk} is a decentralized system that utilizes user-defined private Smart Contracts to automatically implement cryptographic methods. 
Our approach does not require the encoding of custom smart contracts, as it is based on policies stored on chain to encrypt messages.
Bin Li et al.~\cite{RZKPB} introduce RZKPB, a privacy protection system for shared economy 
based on blockchain technology. Similarly to MARTSIA, this approach does not involve third parties and resorts to external data stores.
Differently from their approach, we link data on chain with the data stores so as to permanently store the resource locators. 
Henry et al.~\cite{henry2022random} employ smart contracts that handle 
payment tokens. Banks operate as trustworthy intermediaries to preserve privacy. MARTSIA pursues confidentiality of exchanged information too, although it does not resort to central authorities (the banks) to this end. 
Rahulamathavan et al.~\cite{IoT-ABE} propose a new blockchain architecture for IoT applications that preserves privacy through the use of \ABE. We also utilize ABE in our approach, but 
MARTSIA integrates with existing technologies, whilst their model aims to change the blockchain protocol.
Benhamouda et al.~\cite{BenhamoudaCanAP} introduce a solution that enables a public blockchain to serve as a storage place for confidential data. 
As in our approach, they utilize shared secrets among components. However, their approach discloses the secret when determined conditions are fulfilled, whereas MARTSIA does not reveal secret data on the blockchain. 
In the healthcare domain, Wang et al.~\cite{EHR} create a secure electronic health records system that combines Attribute-Based Encryption, 
Identity-Based Encryption, 
and Identity-Based Signature 
with blockchain technology.
Their architecture is different from ours as the hospital has control over the patient's data and sets the policies, whereas solely the data owners manage data in MARTSIA. 
\ifLong{%
	Pournaghi et al.~\cite{MedSBA}, introduce a system called MedSBA that uses attribute-based encryption to secure medical data on private blockchains. Similarly, However, their approach differs from ours because they use two private blockchains while our approach is based on a public one.
	Nonetheless, their method encrypts the data using AES symmetric cryptography and then encrypts that symmetric key using ABE. We use the same approach because the ABE key would be too short to encrypt long messages. In this way we implement an hybrid solution with symmetric encryption and ABE.%
}
\ifShort{%
	Tran et al.~\cite{SharingSystems} and Pournaghi et al.~\cite{MedSBA} propose approaches for decentralized storage and sharing based on private blockchains. We operate in the context of public blockchains to leverage the higher degree of security given by the general validation of transactions.%
}
Athanere et al.~\cite{ATHANERE20221523} present an approach where the data owner encrypts the file, and then a hashed version of it is stored on a cloud server. The data owner encrypts the data with the public key of the message reader, and the necessary public parameters are generated by an administrator. MARTSIA differs from this solution because it uses \MAABE and symmetric key encryption to encrypt the data instead of a public key. 
\ifLong{%
	Tran et al.~\cite{SharingSystems} propose an approach for decentralized storage and sharing system. In this work, there are different organizations that run a private blockchain and each of them have a worker leader. This leader listens to events in the public blockchain finding transactions for synchronization in the private one. Our work differs from this one because it does not need a private blockchain or a worker leader. %
}
Pham et al.~\cite{B-Box} propose an idea for a decentralized storage system named B-Box, based on \IPFS, \MAABE and blockchains. Though we resort to those building blocks too, we include mechanisms for secure initialization of the authority network, allow for fine-grained access control on message parts, and impede by design any actor from accessing keys. 

\section{Conclusion and future remarks}
\label{sec:conclusion}
In this work, we introduce MARTSIA, a technique that merges blockchain technology with \acrfull{maabe} to regulate data access in the scenario of multi-party business operations. Additionally, our method employs \IPFS for preserving information artifacts, access regulations, and metadata.
We utilize smart contracts to keep the user attributes, establish the access grants to the process participants, and save the connection to IPFS files. MARTSIA allows for a detailed specification of access permissions, ensuring data reliability, persistence, and irrefutability, thus enabling auditability with minimal added costs. 


Our approach exhibits limitations we aim to overcome in future work. If a \DOwner wants to revoke access to data for a particular \Reader, e.g., they can change the policy and encrypt the messages again. However, the old data on \IPFS would still be accessible. Therefore, we are considering the usage of InterPlanetary Name System (IPNS), as it allows for the replacement of existing files. With it, a message can be replaced with a new encryption thereof that impedes {\Reader}s whose grant was revoked to access it. 
\ifLong{%
\ennote{Moreover, although it does not represent a risk from a privacy point of view, everyone at the moment can call MARTSIA's smart contracts functions. To prevent this behavior and restrict unauthorized calls, we plan to implement access management control at the smart contracts level allowing only certified entities to call smart contracts functions. This further strengthens MARTSIA's flow.}%
}
More generally, the life-cycle of data artifacts, policies and smart contracts constitutes a management aspect worth investigating.
From a technological perspective, we are working on the implementation of MARTSIA on other public blockchain platforms such as Algorand%
\footnote{A preliminary version is available at \href{https://github.com/apwbs/MARTSIA-Algorand}{github.com/apwbs/MARTSIA-Algorand}}
~\cite{Chen.Micali/TCS2019:Algorand}
to analyze the benefits and challenges stemming from different \DLTs, including costs.
Also, we are developing an alternative key request protocol for readers that adopts the blockchain as a communication layer so as to avoid direct channels between readers and authorities.
In light of the considerable impact that a correct expression of policies has on the overall approach, we envision automated verification and simulation of policies for future work to properly assist the users in their policy specification task.
Future endeavors also include the integration of Zero Knowledge Proofs~\cite{goldwasser2019knowledge} with \ABE to yield better confidentiality and privacy guarantees, and of decentralized identifiers~\cite{Norta.etal/CS2019:BlockchainEnabledIdentityAuth} and oracles~\cite{Basile.etal/BPMBCF2021:BlockchainProcessesDecentralizedOracles} 
to verify data ownership.
We plan to conduct a formal threat analysis to prove the security of our approach, and run field tests for the empirical evaluation of its robustness.
\subsubsection*{Acknowledgements.}
The work of E.~Marangone and C.~Di~Ciccio was partially funded by the Cyber 4.0 project BRIE, the Sapienza project DRONES, 
by project SERICS (PE00000014) under the NRRP MUR program funded by the EU-NGEU,
and by project PINPOINT (B87G22000450001) under the PRIN MUR program. %
\vspace{-1em}
%
%
\bibliographystyle{splncs04}
\bibliography{bibliography}

\end{document}